\documentclass[11pt]{article}
\newlength{\vshift}
\newlength{\hshift}
\usepackage{amssymb,amsopn}

\newcommand{\newsection}{ \setcounter{equation}{0} \section}
\def\nn{\nonumber }
\def\la{\lambda}
\def\ka{\kappa}

\def\ve{\varepsilon}

\def\Ax{{\mathcal{A}_x}}
\def\Az{{\mathcal{A}_z}}

\def\ds{\stackrel{*}{,}}
\def\x{\hat x}

\def\z{\hat z}
\def\xx{{\hat X}}
\def\aa{{\hat A}}
\def\pat{\partial}

  \newcommand*\ti[5]{{\em #5}, {#1} {\bf #2}, #3 (#4)}
\providecommand{\href}[2]{#2}

\newcommand*\jhep{JHEP}

\newcommand*\pr{Phys. Rev.}

\begin{document}

\begin{titlepage}
\rightline{LMU-TPW 2000-17}
\rightline{MPI-PhT/2000-25}

\vspace{4em}
\begin{center}

{\Large{\bf Enveloping algebra valued gauge transformations for 
non-abelian gauge groups on non-commutative spaces}}

\vskip 3em

{{\bf B.\ Jur\v co${}^{1}$, S.\ Schraml${}^{2,3}$, P.\ Schupp${}^{3}$, J.\ Wess${}^{2,3}$ }}

\vskip 1em

${}^{1}$Max-Planck-Instiut f\"ur Mathematik\\ 
        Vivatgasse 7, D-53111 Bonn\\[1em]

${}^{2}$Max-Planck-Institut f\"ur Physik\\
        F\"ohringer Ring 6, D-80805 M\"unchen\\[1em]

${}^{3}$Sektion Physik, Universit\"at M\"unchen\\
        Theresienstr.\ 37, D-80333 M\"unchen

\end{center}

\vspace{2em}

\begin{abstract} 
An enveloping algebra valued gauge field is constructed, its components 
are functions of the Lie algebra valued gauge field and can be constructed 
with the Seiberg-Witten map. This allows the formulation of a dynamics for 
a finite number of gauge field components on non-commutative spaces. 
\end{abstract}

\vfill

\end{titlepage}\vskip.2cm

\newpage
\setcounter{page}{1}

%%%%%%%%%%%%%%%%%%%%%%%%%%%%%%%%%%%%%%%%%%%%%%%%%%%%%%%%%%%%%%%%%%%%%%%%%%%%%%%%%%%%%%%%%%%%
\newsection{Introduction}

Gauge theories on non-commutative spaces~\cite{MSSW,CDS,SW} cannot be formulated with Lie algebra valued
infinitesimal transformations and  consequently not with Lie algebra valued
gauge fields. In the composition of the infinitesimal transformations
commutators and anticommutators of the generators of the gauge group
will appear, eventually generating all the higher powers of 
the generators. Thus the enveloping algebra of the Lie algebra
seems to be a proper setting for such a gauge theory. This,
however, is not very attractive because the enveloping algebra
is infinite-dimensional, requiring an infinite number of 
coordinate dependent transformation parameters and an infinite
number of gauge fields as a consequence. 

In this paper we show that enveloping algebra valued 
infinitesimal transformations as well as  enveloping
algebra valued gauge fields can be restricted such that they
depend on the Lie algebra valued parameters and the Lie 
algebra valued gauge fields and their space-time derivatives only.
This renders the number of independent parameters and
gauge fields to be the same as for the Lie algebra valued
gauge theories. The coefficient functions of all the higher powers
of the generators of the gauge group are functions of the
coefficients of the first power. The construction of the 
dependent coefficients is based on the Seiberg-Witten map~\cite{SW}.
The existence of this map can be proven in general~\cite{JS,JSW,JSW2}, here we
demonstrate this map by explicitely calculating the expansion to first order
in a parameter that characterizes the deviation from commuting coordinates.

As a method we use the $*$-product formulation of the
algebra~\cite{Weyl,Wigner,Moyal,BFFLS,Kontsevich,Sternheimer}. The objects are
functions of commuting variables, the algebraic  non-commutative properties
are encoded in the $*$-product. In the following chapter we introduce this
formalism, it can be used for all algebras that have the
Poincare-Birkhoff-Witt property. In this paper we restrict the algebra to the
algebra of the non-commuting  coordinates and the Lie algebra of the gauge
group. 

For non-commuting spaces the concept of a gauge theory
can already be introduced by defining covariant coordinates
without speaking about derivatives~\cite{MSSW}. 
In general the algebraic setting of the theory will require
an extension of the algebra by derivatives. This formalism has
been developed for quantum planes~\cite{WZ}. 

For the canonical structure
of the non-commuting coordinates it is shown in this paper that
the derivatives can be obtained from the coordinates. Derivatives
do not have to be introduced separately. For several examples of
quantum planes this is true as well~\cite{W}.

For the canonical structure integration can be defined as well. This
is shown in chapter~\ref{dynamics}. This allows us to formulate the dynamics with 
an action. A gauge invariant action for the gauge field can be constructed
from the gauge covariant tensors that agrees with the usual 
gauge invariant action in the limit of commuting spaces.

The approach generalizes to other non-commutative spaces and it is in particular
possible to choose another non-commutative internal space instead of the
Lie algebra. For an internal canonical structure this scenario has been studied
in~\cite{Wise}.

%%%%%%%%%%%%%%%%%%%%%%%%%%%%%%%%%%%%%%%%%%%%%%%%%%%%%%%%%%%%%%%%%%%%%%%%%%%%%%%%%%%%%%%%
\newpage
\newsection{Non-commutative Spaces and the $*$-Product Formalism}

The coordinates $\z^i$, $(i=1,\ldots N)$ of a non-commutative space structure are
subject to relations. We have in mind the relations for a 
canonical structure
\begin{equation}
\label{can-struct}
  [\z^i,\z^j]=i\theta^{ij},\qquad \theta^{ij}\in\mathbb{C},
\end{equation}
for a Lie structure
\begin{equation}
  [\z^i,\z^j]=if^{ij}{}_k\z^k,\qquad f^{ij}{}_k\in\mathbb{C},
\end{equation}
or for a quantum plane structure
\begin{equation}
  [\z^i,\z^j]=iC^{ij}_{kl}\z^k\z^l, \qquad C^{ij}_{kl}\in\mathbb{C}.
\end{equation}
The non-commutative space can be defined as the associative 
algebra over $\mathbb{C}$, which consists of the algebra freely 
generated by the coordinates and then divided by the ideal ${\cal R}$ 
generated by the relations:
\begin{equation}
  \Az=\frac{\mathbb{C}[[\z^1,\ldots, \z^N]]}{\cal R}.
\end{equation}
Formal power series are accepted. Among these algebras
we will restrict our attention to those that have a basis
with the Poincare-Birkhoff-Witt property (PBW). This means that
when considered as a graded algebra the subspace of 
polynomials of fixed degree has the same dimension as the 
corresponding subspace of the polynomials of commuting variables.
In this case any element of $\Az$ is defined by its coefficient
function and vice versa.
\begin{eqnarray}
  \label{dia-prod}
  &&\hat{f}=\sum_{L=0}^{\infty}f_{i_1,\ldots, i_L}:\z^{i_1}\cdots \z^{i_L}:\nn\\
  &&\hat{f}\sim\{f_{i}\}.
\end{eqnarray}
$:\z^{i_1}\cdots \z^{i_L}:$ denotes an element of the basis defined 
by some prescribed ordering, e.g., normal order $i_1\le i_2\ldots\le i_L$ or,
e.g., totaly symmetric.   The product of two elements will have its own
coefficient function, this defines the diamond product
\begin{equation}
  \hat{f}\hat{g}=\hat{h}\quad\sim\quad\{f_i\}\diamond\{g_i\}=\{h_i\}.
\end{equation}
The algebraic properties are now all encoded in the
$\diamond$ product. 

Next we associate a function $f$ of commuting variables with 
an element of the algebra, say $\hat{f}$, by substituting 
the commuting variable $z^1,\ldots,z^N$ for the non-commuting variables
in (\ref{dia-prod}) 
\begin{equation}
  \hat{f}=\sum f_{i_1\ldots i_L}:\z^{i_1}\cdots\z^{i_L}:\quad\sim\quad f(z)=\sum f_{i_1\ldots i_L}z^{i_1}\cdots z^{i_L}
\end{equation}
The diamond product leads to a bilinear $*$-product of functions:
\begin{equation}
  \{f_i\}\diamond\{g_i\}=\{h_i\}\quad\sim\quad (f*g)(z)=h(z).
\end{equation}
This star product has been discussed in reference~\cite{MSSW}.

For the canonical structure it is the Moyal-Weyl
product~\cite{Weyl,Wigner,Moyal}: 
\begin{equation}
\label{MoWe-prod}
  f*g=\left.e^{\frac{i}{2}\frac{\pat}{\pat z^i}\theta^{ij}\frac{\pat}{\pat z'^j}}f(z)g(z')\right|_{z'\to z}.
\end{equation}

For the Lie structure we have:
\begin{equation}
  f*g=\left.e^{\frac{i}{2}x^lg_l(i\frac{\pat}{\pat z'},i\frac{\pat}{\pat z''})}f(z')g(z'')\right|_{z'\to z\atop z''\to z},
\end{equation}
where $g_l$ is defined by group multiplication:
\begin{equation}
  e^{ik_l\z^l}e^{ip_l\z^l}=e^{i\{k_l+p_l+\frac{1}{2}g_l(k,p)\}\z^l}.
\end{equation}
The first terms are easily calculated from the
Baker-Campbell-Hausdorff formula:
\begin{eqnarray}
\label{BCH}
  e^Ae^B&=&e^{A+B+\frac{1}{2}[A,B]+\frac{1}{12}([A,[A,B]]+[B,[B,A]])+\ldots}\nn\\
  g_l(k,p)&=&-k_ip_jf^{ij}{}_l+\frac{1}{6}k_ip_j(p_k-k_k)f^{ij}{}_nf^{nk}{}_l+\ldots.
\end{eqnarray}

For the quantum plane structure we have as an example the $*$-product for the 
Manin plane:
\begin{eqnarray}
  xy&=&qyx\nn\\
  f*g &=& \left. q^{\frac{1}{2}(-x'\frac{\pat}{\pat x'}y
\frac{\pat}{\pat y}+x\frac{\pat}{\pat x}y'\frac{\pat}{\pat y'})}f(x,y)g(x',y')
\right|_{x'\to x \atop y'\to y}
\end{eqnarray}

%%%%%%%%%%%%%%%%%%%%%%%%%%%%%%%%%%%%%%%%%%%%%%%%%%%%%%%%%%%%%%%%%%%%%%%%%%%%%%%%%%%%%%%%
\newsection{Enveloping algebra valued connection}

A non-abelian gauge theory on a non-commutative space carries
two algebraic structures, the algebra $\Ax$ discussed
above and the non-abelian Lie algebra ${\cal A}_T$ of the gauge group 
with the generators $T^1,\ldots,T^M$ and the relations:
\begin{equation}
  [T^a,T^b]=if^{ab}{}_cT^c.
\end{equation}
It is natural to treat both algebras on the same footing and
to denote the generating elements of the big algebra by $\z^i$:
\begin{eqnarray}
  \z^i&=&\{\x^1,\ldots,\x^N,T^1,\ldots,T^M\}\nn\\
  \Az&=&\frac{\mathbb{C}[[\z^1,\ldots, \z^{N+M}]]}{\cal R}.
\end{eqnarray}
The $*$-product formalism as developed in the previous chapter
can now be applied to the algebra $\Az$ as well.

We study functions of the commuting variables $x^{\nu}$, $(\nu=1,\ldots, N)$ and
$t^a$, $(a=1,\ldots,M)$ and define the star product reflecting the algebraic
properties of the algebra~$\Az$.

In the case of a canonical structure for the space variables
$x^{\nu}$ we have
\begin{eqnarray} \label{blub}
  \lefteqn{(F*G)(z)=}\nn\\
  &&e^{\frac{i}{2}\left(\theta^{\mu\nu}
  \frac{\pat}{\pat x'^{\mu}}
  \frac{\pat}{\pat x''^{\nu}}+t^ag_a(i
  \frac{\pat}{\pat t'},i\frac{\pat}{\pat t''})\right)}\nn\\
  &&\times F(x',t')G(x'',t'')\Big|_{x'\to x, x''\to x\atop t'\to t, t''\to t}. 
\end{eqnarray}
To exemplify the formalism we shall concentrate on this structure in what follows. 

To define gauge theories we first define fields. These are
elements of the algebra $\Ax$ that form a representation of the
$T$-algebra. Under a  gauge transformation they transform as
follows:
\begin{equation}
\label{gaugetrafo}
  \delta\hat{\psi}=i\hat{\alpha}\hat{\psi},\quad \hat{\psi}\in\Ax,\quad \hat{\alpha}\in\Az.
\end{equation}
The action of the generators $T$ of the Lie algebra on 
$\hat{\psi}$ is defined as $\hat{\psi}$ is supposed to form a representation of ${\cal A}_T$.
Thus $\delta\hat{\psi}\in\Ax$ despite $\hat{\alpha}\in\Az$.

Independent of a representation we have defined $\hat{\alpha}$
as an element of the enveloping algebra of the gauge group
and not as Lie algebra-valued, as we would have done it for
commuting spaces. We say $\hat{\alpha}$ is enveloping algebra-valued.
The same will be true for the connection that we introduce to
define covariant coordinates \cite{MSSW}
\begin{equation}
  \xx^{\nu}=\x^{\nu}+\aa^{\nu},\qquad \aa^{\nu}\in \Az.
\end{equation}
We demand that $\hat{X}^{\nu}\hat{\psi}$ transforms covariantly:
\begin{equation}
  \delta\xx^{\nu}\hat{\psi}=i\hat{\alpha}\xx^{\nu}\hat{\psi}
\end{equation}
and find that this defines the transformation law of the 
enveloping algebra-valued connection $\aa^{\nu}$:
\begin{eqnarray}
  &&\delta\aa^{\nu}=-i[\x^{\nu},\hat{\alpha}]+i[\hat{\alpha},\aa^{\nu}],\nn\\
  &&\aa^{\nu}\in\Az,\quad \hat{\alpha}\in\Az,\quad \delta\hat{A}^{\nu}\in\Az.
\end{eqnarray}
At first sight it seems that an enveloping algebra-valued 
connection has infinitely many component fields and thus is
not very useful. We shall show, however, that all the 
component fields can be obtained from a Lie algebra-valued
connection by a Seiberg-Witten map \cite{SW,JS,JSW2}. This was also
observed in~\cite{Bonora}, where a result in this direction has been obtained
for SO(n) and Sp(n).  To show this
we cast the algebraic setting into the $*$-product formalism.
The transformation of the connection is then
\begin{equation}
  \delta A^{\nu}=-i[x^{\nu}\ds{\alpha}]+i[{\alpha}\ds A^{\nu}].
\end{equation}
We treat the canonical case in more detail, the $*$-product
in this case is given in Equation (\ref{blub}).

For the first term in the variation of $A^{\nu}$ we obtain
\begin{equation}
\label{x-alpha-kommu}
  -i[x^{\nu}\ds\alpha]=\theta^{\nu\rho}\frac{\pat}{\pat x^{\rho}}\alpha.
\end{equation}
The variation of $A^{\nu}$ itself starts with a linear term in $\theta$, 
we therefore assume, as in reference \cite{MSSW}, that $A^{\nu}$ starts with a 
linear term in $\theta$ as well:
\begin{eqnarray}
\label{delta-v}
  A^{\nu}&=&\theta^{\nu\rho}V_{\rho}\nn\\
  \delta V_{\rho}&=&\frac{\pat}{\pat x^{\rho}}\alpha+i[\alpha\ds V_{\rho}].
\end{eqnarray}
As in reference \cite{MSSW} we expand in $\theta$, but not in $g_a$:
\begin{eqnarray}
  \label{xt-star}
  &&f*g=\left.\left\{1+\frac{i}{2}\frac{\pat}{\pat x^{\nu}}\theta^{\nu\mu}\frac{\pat}{\pat x'^{\mu}}+\ldots\right\}f(x,t')\circledast g(x',t'')\right|_{x'\to x\atop t'\to t, t''\to t}\nn\\
  &&f(x,t')\circledast g(x',t'')=e^{\frac{i}{2}t^ag_a(i\frac{\pat}{\pat t'},i\frac{\pat}{\pat t''})}f(x,t')g(x',t'').
\end{eqnarray}
We first treat Equation (\ref{delta-v}) to zeroth order in $\theta$ 
and show that it can be solved by assuming $V$ and $\alpha$
to be linear in $t$. This has to be expected as $\theta=0$
corresponds to the usual gauge theory on commuting spaces
where the infinitesimal transformation and the connection
are Lie algebra-valued. To zeroth order:
\begin{eqnarray}
  \alpha&=&\alpha^{1}_at^a,\nn\\
  V_{\rho}&=&a^{1}_{\rho,a}t^a.
\end{eqnarray}
From (\ref{delta-v}) we obtain, as expected~\cite{JSW2}:
\begin{equation}
\label{delta-a}
  \delta a^1_{\rho,a}=\frac{\pat \alpha^1_a}{\pat x^{\rho}}-f^{bc}{}_a\alpha^{1}_ba^{1}_{\rho,c}.
\end{equation}

We turn to first order in $\theta$ in the variation of $V_{\rho}$, Equation (\ref{delta-v}). 
The contributions that come
from the zero order terms of $\alpha$ and $V_{\rho}$ are at most of 
second order in $t$. This is the case because 
$t^ag_a(i\frac{\pat}{\pat t'},i\frac{\pat}{\pat t''})$ reduces 
the power of $t$ by at least one. 
The terms of order zero in $g_a$ actually contribute exactly in order $t^2$.
Their contribution to $\delta V_{\rho}$ is
\begin{equation}
  \delta V_{\rho}=\theta^{\nu\mu}\pat_{\nu}\alpha_a^1\pat_{\mu}a^1_{\rho,b}t^at^b+\ldots.
\end{equation}
If we now assume that the terms of $\alpha$ and $V_{\rho}$ linear  
in $\theta$ are all of second order in $t$ we get a consistent
set of equations. We define to first order in $\theta$: 
\begin{eqnarray}
  \alpha&=&\alpha^{1}_at^a+\alpha^{2}_{ab}t^at^b+\ldots\nn\\
  V_{\rho}&=&a^{1}_{\rho,a}t^a+a^{2}_{\rho,ab}t^at^b+\ldots
\end{eqnarray}
$\alpha^{1}$ and $a^{1}$ are of order zero in $\theta$ and $\alpha^{2}$ 
and $a^{2}$ of first order. This expansion in $t$ leads to an
expansion in $g_a$ of the $\circledast$-product, because higher order $t$-derivatives vanish.

In the calculation that now follows we have to
use $g_a$ to the order given in (\ref{BCH}). The term with
three derivatives, however, vanishes on commutators of $*$-products 
because it is symmetric under the exchange 
of $k$ and $p$.

The result of the calculation is:
\begin{eqnarray}
\label{bestim-gl}
  \delta a^{2}_{\rho,ab}t^at^b&=&\pat_{\rho}\alpha^{2}_{ab}t^at^b\nn\\
  &&-\theta^{\nu\mu}\pat_{\nu}\alpha^{1}_{a}\pat_{\mu}a^{1}_{\rho,b}t^at^b\nn\\
  &&-2f^{bc}{}_a\left\{\alpha^{1}_{b}a^{2}_{\rho,cd}+\alpha^{2}_{bd}a^{1}_{\rho,c}\right\}t^dt^a.
\end{eqnarray}
This can be brought closer to the form of reference \cite{MSSW}.
We introduce the Lie algebra valued $\ve$ and the enveloping algebra-valued
$G_{\rho}$:  \begin{equation}
\label{eps-G}
  \ve=\alpha^{1}_bT^b,\quad G_{\rho}=a^{2}_{\rho,cd}T^cT^d
\end{equation}
and compute the commutator 
\begin{equation}
\label{eps-G-Kommu}
  i[\ve,G_{\rho}]=-\alpha^{1}_ba^2_{\rho,cd}f^{bc}{}_l\{T^lT^d+T^dT^l\}.
\end{equation}
This is true because $a^{2}_{\rho,cd}$ is symmetric in $c$ and $d$. Now 
we remember that we have used a star product that corresponds
to a completely symmetrical version of the monomials of the 
bases. Thus we have to replace 
\begin{equation} 
  \frac{1}{2}\{T^lT^d+T^dT^l\}\quad\sim\quad t^lt^d
\end{equation}
and obtain from (\ref{eps-G-Kommu}) the corresponding term in (\ref{bestim-gl}). The other
term derives from the commutator 
\begin{eqnarray}
  &&i[\gamma,a_{\rho}],\quad\mbox{with}\nn\\
  &&\gamma=\alpha^{2}_{bd}T^bT^d,\quad a_{\rho}=a^1_{\rho,l}T^l.
\end{eqnarray}
This is exactly the structure as in reference \cite{MSSW}:
\begin{eqnarray}
  \delta G_{\nu} &=& \pat_{\nu}\gamma 
- \frac{1}{2}\theta^{\ka\la}\left\{\pat_{\ka} \ve , \pat_{\la} a_{\nu}\right\}\nn\\
  &&+ i[\ve,G_{\nu}]+i[\gamma,a_{\nu}].
\end{eqnarray}                                

It has the solution already found in \cite{MSSW,SW}.
\begin{eqnarray}
\label{solution}
  \alpha^2_{ab}t^at^b&=&\frac{1}{2}\theta^{\nu\mu}\pat_{\nu}\alpha^1_aa^1_{\mu,b}t^at^b\nn\\
  a^2_{\rho,ab}t^at^b&=&-\frac{1}{2}\theta^{\nu\mu}a^1_{\nu,a}(\pat_{\mu}a_{\rho,b}^1+F^1_{\mu\rho,b})t^at^b,
\end{eqnarray}
where $F_{\nu\rho,b}^1=\pat_{\nu}a_{\mu,b}^1-\pat_{\mu}a_{\nu,b}^1+f^{cd}{}_{b}a_{\nu,c}^1a_{\mu,d}^1$.

The procedure can now be generalized to all the higher 
powers of $\theta$. We always assume that $\alpha^{n}$ and $a_{\rho}^{n}$ are of 
power $\theta^{n-1}$ and polynomials in $t$ of degree $n$. This  
amounts to a power series expansion in $g_a$ as well and the existence 
of the Seiberg-Witten map to all orders follows from the existence
of the Seiberg-Witten map in the general setting of the 
previous chapter. This is discussed in detail in reference~\cite{JSW2},
where the full non-abelian Seiberg-Witten map is constructed.

It is straightforward to generalize this formalism to a $*$-product, with a 
coordinate dependent $\theta$ underlying the non-commutative, e.g. Lie or quantum 
plane, structure. 
We give the result:
\begin{eqnarray}
  \alpha&=&\alpha^1_at^a+\frac{1}{2}\theta^{\nu\mu}\pat_{\nu}\alpha^1_aa^1_{\mu,b}t^at^b\nn\\
  A^{\nu}&=&\theta^{\nu\mu}a^1_{\mu,a}t^a\nn\\
  &&-\frac{1}{2}\theta^{\sigma\mu}a^1_{\sigma,a}\left(\pat_{\mu}(\theta^{\nu\rho}a_{\rho,b}^1)+\theta^{\nu\rho}F^1_{\mu\rho,b}\right)t^at^b
\end{eqnarray}
For the Lie structure:
\begin{equation}
  \theta^{\mu\nu}=f^{\mu\nu}{}_{\ka}x^{\ka}.
\end{equation}
For the quantum plane:
\begin{equation}
  \theta^{\mu\nu}=-ihxy, \qquad h=\ln q.
\end{equation}

Let us consider once more the gauge transformations (\ref{gaugetrafo}). 
We see that $\alpha$ is not an arbitrary element of the algebra $\Az$.
There are only $M$ (dimension of the Lie group to be gauged)
free parameters $\alpha^{1}_l(x)$, all the higher-order terms in the enveloping 
algebra can be expressed in terms of these parameters and the 
gauge field $a^{1}_{\rho,l}$ and their derivatives. This can be achieved by the
Seiberg-Witten map.

To summarize, the Lie algebra-valued term $\ve=\alpha_a^1(x)T^a$ of $\alpha$  
determines all the other terms in the enveloping algebra: 
\begin{equation}
\label{alph-env}
  \alpha=\alpha^1_aT^a+\frac{1}{4}\theta^{\nu\mu}\pat_{\nu}\alpha^1_aa^1_{\mu,b}(T^aT^b+T^bT^a)+\ldots
\end{equation}
Thus a gauge transformation is determined by $\alpha^1(x)$ and $a^1_{\rho}(x)$, it is defined by:
\begin{equation}
\label{res-gauge-trafo}
  \delta_{\alpha^1}\psi=i\alpha(\alpha^1,a^1_{\rho})*\psi.
\end{equation}
The composition of two transformations is defined for arbitrary 
enveloping algebra-valued transformations as follows:
\begin{equation}
  \delta_{\alpha}\delta_{\beta}-\delta_{\beta}\delta_{\alpha}=\delta_{i(\beta*\alpha-\alpha*\beta)}.
\end{equation}
We show that this is also true for the restricted transformation defined by $\alpha^1$:
\begin{equation}
\label{alph1-komm}
  \delta_{\alpha^1}\delta_{\beta^1}-\delta_{\beta^1}\delta_{\alpha^1}=\delta_{i(\beta*\alpha-\alpha*\beta)^1}.
\end{equation}
This reflects the compositon of standard Lie algebra-valued gauge transformations. 
We show this to first order in $\theta$, using the $*$-formalism.
\begin{eqnarray}
\label{alpha+beta}
  \alpha&=&\alpha^1_at^a+\frac{1}{2}\theta^{\nu\mu}\pat_{\nu}\alpha^1_aa^1_{\mu,b}t^at^b\nn\\
  \beta&=&\beta^1_at^a+\frac{1}{2}\theta^{\nu\mu}\pat_{\nu}\beta^1_aa^1_{\mu,b}t^at^b
\end{eqnarray}
We compute $[\alpha\ds\beta]$ to first order in $\theta$ from Equation (\ref{xt-star}):
\begin{eqnarray}
  [\alpha\ds\beta]&=&i\alpha_a^1\beta^1_bf^{ab}{}_ct^c\nn\\
  &&+\theta^{\nu\mu}\Big\{\frac{i}{2}\pat_{\nu}(\alpha^1_a\beta^1_bf^{ab}{}_{d})a_{\mu,c}\nn\\
  &&\qquad+\frac{i}{2}(\alpha^1_a\pat_{\nu}\beta^1_d-\beta_a^1\pat_{\nu}\alpha_d^1)a^1_{\mu,b}f^{ab}{}_{c}\nn\\
  &&\qquad+i\pat_{\nu}\alpha^1_d\pat_{\mu}\beta^1_c\Big\}t^dt^c.
\end{eqnarray}
Now we compute from (\ref{gaugetrafo}) 
\begin{equation}
  (\delta_{\beta^1}\delta_{\alpha^1}-\delta_{\alpha^1}\delta_{\beta^1})\hat{\psi}=-(\hat{\alpha}\hat{\beta}-\hat{\beta}\hat{\alpha})\hat{\psi}+i((\delta_{\beta^1}\hat{\alpha})-(\delta_{\alpha^1}\hat{\beta}))\hat{\psi}.
\end{equation}
The second term arises because $\alpha$ depends on the gauge fields $a^1_{\mu}$ that 
transforms under gauge transformations 
\begin{equation}
\label{delta-alpha}
  \delta_{\beta^1}\alpha=\frac{1}{2}\theta^{\rho\sigma}\pat_{\rho}\alpha^1_a(\pat_{\sigma}\beta^1_b-f^{cd}{}_{b}\beta^1_ca^1_{\sigma,d})t^dt^b.
\end{equation}
We are now ready to compute $\delta_{\beta^1}\delta_{\alpha^1}-\delta_{\alpha^1}\delta_{\beta^1}$ 
and obtain
\begin{eqnarray}
  \lefteqn{\delta_{\beta^1}\delta_{\alpha^1}-\delta_{\alpha^1}\delta_{\beta^1}=i(\delta_{\beta^1}\alpha-\delta_{\alpha^1}\beta)-[\alpha\ds\beta]}\nn\\
  &=& -\left(i\alpha_a^1\beta^1_bf^{ab}{}_ct^c+\frac{1}{2}\theta^{\nu\mu}\pat_{\nu}(i\alpha^1_a\beta^1_bf^{ab}{}_d)a^1_{\mu,c}t^dt^c\right).
\end{eqnarray}
This is exactly the formula (\ref{alph1-komm}) that we obtain if we start from 
the Lie algebra-valued part of $[\alpha\ds\beta]$
\begin{equation}
  [\alpha\ds\beta]=i\alpha^1_a\beta^1_bf^{ab}{}_cT^c+\ldots
\end{equation}
This shows that our restricted enveloping algebra-valued form of the parameters  
is respected by the commutator of two transformations.

In reference \cite{MSSW} we have introduced tensors
\begin{equation}
  \hat{T}^{\mu\nu}=[\hat{X}^{\mu},\hat{X}^{\nu}]-i\hat{\theta}^{\mu\nu},
\end{equation}
that transform
\begin{equation}
\label{T-trafo}
  \delta\hat{T}^{\mu\nu}=i[\hat{\alpha},\hat{T}^{\mu\nu}]
\end{equation}
under the general enveloping algebra valued gauge transformation. 
$\hat{\theta}^{\mu\nu}$ is the respective term for all the three structures, 
canonical, Lie and quantum plane. Equation (\ref{T-trafo}) will also be true for 
our restricted gauge transformation (\ref{res-gauge-trafo}). 

For the canonical case to exemplify:
\begin{eqnarray}
  T^{\mu\nu}&=&i\theta^{\mu\ka}\pat_{\ka}A^{\nu}-i\theta^{\nu\la}\pat_{\la}A^{\mu}+A^{\mu}*A^{\nu}-A^{\nu}*A^{\mu}\nn\\
  &=&\theta^{\mu\ka}\theta^{\nu\la}\left\{\pat_{\ka}V_{\la}-\pat_{\la}V_{\ka}+V_{\ka}*V_{\la}-V_{\la}*V_{\ka}\right\}.
\end{eqnarray}
It is natural to introduce the field strength
\begin{equation}
\label{field-str}
  F_{\ka\la}=\pat_{\ka}V_{\la}-\pat_{\la}V_{\ka}+V_{\ka}*V_{\la}-V_{\la}*V_{\ka}.
\end{equation}
It is easy to compute the first order correction to the classical field strength 
$F^1$ defined after Equation (\ref{solution}):
\begin{eqnarray}
\label{field-str-correction}
  F_{\ka\la,a}t^a&=&F^1_{\ka\la,a}t^a\nn\\
  &&+\theta^{\mu\nu}\Big\{F^1_{\ka\mu,a}F^1_{\la\nu,b}\nn\\
  &&\quad-\frac{1}{2}a^1_{\mu,a}\left((D_{\nu}F^1_{\ka\la})_b+\pat_{\nu}F^1_{\ka\la,b}\right)\Big\}t^at^b.
\end{eqnarray}
In this formula a covariant derivative of $F^1$ is used: $(D_{\nu}F^1_{\ka\la})_b=\pat_{\nu}F^1_{\ka\la,b}+a_{\nu,c}F^1_{\ka\la,d}f^{cd}{}_b$.
This expression can also be obtained from reference \cite{SW}. 

Using the transformation law for $\alpha$ (\ref{delta-alpha}) and $a^1$ (\ref{delta-a}) 
we find as expected 
\begin{equation}
  \delta_{\alpha^1}F_{\ka\la}=i[\alpha\ds F_{\ka\la}], 
\end{equation}
with the restricted form of $\alpha$.

%%%%%%%%%%%%%%%%%%%%%%%%%%%%%%%%%%%%%%%%%%%%%%%%%%%%%%%%%%%%%%%%%%%%%%%%%%%%%%%%%
\newsection{Gauge covariant dynamics}
\label{dynamics}

The $*$-formalism can be used to formulate a dynamics on non-commutative spaces. 
The coefficient functions $f(x)$ are the objects for which dynamical laws can be 
defined. In general we have to enlarge the algebra by derivatives for this purpose. 
For the quantum plane structure such derivatives have been introduced in \cite{WZ} 
in a purely algebraic approach. For the algebra extended by derivatives the formalism 
developed in the second chapter can be used. 

For the canonical structure we can define derivates following the same strategy as for 
the quantum plane structure. Derivatives have to be defined in such a way, 
that they do not lead to new relations for the coordinates. 
Proceeding this way we can define a Leibniz rule:
\begin{equation}
  \hat{\pat}_{\mu}\x^{\nu}=\delta^{\nu}_{\mu}+d^{\nu\rho}_{\mu\sigma}\x^{\sigma}\hat{\pat}_{\rho},
\end{equation}
where the coefficients $d^{\nu\rho}_{\mu\sigma}\in\mathbb{C}$ have to be 
choosen in such a way that
\begin{eqnarray}
  \lefteqn{\hat{\pat}_{\rho}\left\{[\x^{\mu},\x^{\nu}]-i\theta^{\mu\nu}\right\}=}\nn\\
  &=& \delta^{\mu}_{\rho}\x^{\nu}-\delta^{\nu}_{\rho}\x^{\mu}+d^{\mu\nu}_{\rho\ka}\x^{\ka}-d^{\nu\mu}_{\rho\ka}\x^{\ka}\nn\\
  &&+\x^{\ka}\x^{\beta}\left\{d^{\mu\sigma}_{\rho\ka}d^{\nu\alpha}_{\sigma\beta}-d^{\nu\sigma}_{\rho\ka}d^{\mu\alpha}_{\sigma\rho}\right\}\hat{\pat}_{\alpha}\nn\\
  &&-i\theta^{\mu\nu}\hat{\pat}_{\rho}
\end{eqnarray}
does not lead to new relations when $\hat{\pat}$ is brought to the right hand side. 
This is the case if we define
\begin{equation}
  d^{\nu\rho}_{\mu\sigma}=\delta^{\nu}_{\sigma}\delta^{\rho}_{\mu},
\end{equation}
or simply
\begin{equation}
  \hat{\pat}_{\rho}\x^{\mu}=\delta^{\mu}_{\rho}+\x^{\mu}\hat{\pat}_{\rho}.
\end{equation}
If we now compare this with (\ref{can-struct}) we see that 
\begin{equation}
  \x^{\alpha}-i\theta^{\alpha\rho}\hat{\pat}_{\rho} 
\end{equation}
commutes with all coordinates. This allows us to divide the algebra by 
the ideal generated by the relation (see also~\cite{Ho,AG})
\begin{equation}
  \x^{\alpha}-i\theta^{\alpha\rho}\hat{\pat}_{\rho}=0.
\end{equation}
The star product, known for the coordinates is now defined for the 
derivations as well
\begin{eqnarray}
\label{star-deriv}
  \pat_{\rho}*f&=&-i\theta^{-1}_{\rho\sigma}x^{\sigma}*f\nn\\
  &=& -i\theta^{-1}_{\rho\sigma}\left([x^{\sigma}\ds f]+f*x^{\sigma}\right)\nn\\
  &=&\frac{\pat}{\pat x^{\rho}}f+f*\pat_{\rho}.
\end{eqnarray}
We have used (\ref{x-alpha-kommu}).

%Covariant derivations follow:
%\begin{equation}
%  \hat{\pat}_{\rho}+\hat{V}_{\rho}=-i\theta^{-1}_{\rho\sigma}\x^{\sigma}.
%\end{equation}
%This is the reason why $\hat{V}_{\rho}$ defind in (\ref{delta-v}) enters the 
%field strength. 

For the canonical structure an integral can be defined:
\begin{equation}
  \int \hat{f}=\int d^Nx\;f(x^1,\ldots x^N).
\end{equation}
For the moment it is simpler to consider just functions of $x^i$ that do not 
depend on the variables $t^a$ as well. The $*$-product simply is the one of (\ref{MoWe-prod}): 
\begin{eqnarray}
  f*g&=&\left.e^{\frac{i}{2}\frac{\pat}{\pat x^i}\theta^{ij}\frac{\pat}{\pat x'^j}}f(x)g(x')\right|_{x'\to x}\nn\\
  &=&\int d^Nx'\;\delta(\vec{x}-\vec{x}')e^{\frac{i}{2}\frac{\pat}{\pat
x^i}\theta^{ij}\frac{\pat}{\pat x'^j}}f(x)g(x') \nn\\
  &=&.   \end{eqnarray}
$\delta(\vec{x}-\vec{x}')$ is the product of $N$ $\delta$-functions:
\begin{equation}
  \delta(\vec{x}-\vec{x}')=\delta(x^1-x'^1)\cdots\delta(x^N-x'^N).
\end{equation}
We now show that
\begin{equation}
  \int \hat{f}\hat{g}=\int\hat{g}\hat{f}.
\end{equation}
We use the $*$-formalism:
\begin{equation}
  \int \hat{f}\hat{g}=\int d^Nxd^Nx'\;\delta(\vec{x}-\vec{x}')e^{\frac{i}{2}\frac{\pat}{\pat x^i}\theta^{ij}\frac{\pat}{\pat x'^j}}f(x)g(x').
\end{equation}
Partial integration:
\begin{equation}
  \int \hat{f}\hat{g}=\int d^Nxd^Nx'\;f(x)g(x')e^{\frac{i}{2}\frac{\pat}{\pat x^i}\theta^{ij}\frac{\pat}{\pat x'^j}}\delta(\vec{x}-\vec{x}').  
\end{equation}
The $\delta$-function depends on $\vec{x}-\vec{x}'$. Thus
\begin{equation}
  \frac{\pat}{\pat x^l}\delta(\vec{x}-\vec{x}')=-\frac{\pat}{\pat x'^l}\delta(\vec{x}-\vec{x}').
\end{equation}
This leads to
\begin{equation}
  \int \hat{f}\hat{g}=\int d^Nxd^Nx'\;f(x)g(x')e^{\frac{i}{2}\frac{\pat}{\pat x'^i}\theta^{ij}\frac{\pat}{\pat x^j}}\delta(\vec{x}-\vec{x}').  
\end{equation}
Partial integration again:
\begin{eqnarray}
  \int \hat{f}\hat{g}&=&\int d^Nxd^Nx'\;\delta(\vec{x}-\vec{x}')e^{\frac{i}{2}\frac{\pat}{\pat x'^i}\theta^{ij}\frac{\pat}{\pat x^j}}g(x')f(x)\nn\\
  &=&\int \hat{g}\hat{f}.
\end{eqnarray}
We also find Stokes theorem
\begin{equation}
\label{stokes}
  \int [\hat{\pat}_l,\hat{f}]=\int d^Nx\;[\pat_l\ds f]=\int d^Nx\frac{\pat}{\pat x^l}f=0
\end{equation}
from (\ref{star-deriv}) for functions that vanish at the boundary. 
Partial integration of the $*$-derivative follows now from (\ref{stokes}) and the Leibniz rule 
\begin{equation}
  [\pat_l\ds (f*g)]=([\pat_l\ds f])*g+f*([\pat_l\ds g]),
\end{equation}
which is true, because $\theta$ is $x$ independent. 

With this integral we can define an action. A tensor that transforms as in (\ref{T-trafo}): 
\begin{equation}
  \delta\hat{L}=i[\hat{\alpha},\hat{L}]
\end{equation}
will lead to a gauge invariant action:
\begin{equation}
  W=\int d^Nx\; \mbox{Tr}\hat{L},\qquad \delta W=0.
\end{equation}
The trace has to be taken for the group generators. A proper action would be 
\begin{equation}
  L=\frac{1}{4}F_{\ka\la}F^{\ka\la},
\end{equation}
where $F_{\ka\la}$ has been defined in (\ref{field-str}). 
The first correction term in $\theta$ to the classical field strength has been 
computed in (\ref{field-str-correction}). 

We have thus formulated dynamics on a non-commutative space entirely whithin the
standard framework of quantum field theory. The method can be extended to the
treatment of matter fields aswell.

%%%%%%%%%%%%%%%%%%%%%%%%%%%%%%%%%%%%%%%%%%%%%%%%%%%%%%%%%%%%%%%%%%%%%%%%%%%%%%%%%

%%%%%%%%%%%%%%%%%%%%%%%%%%%%%%%%%%%%%%%%%%%%%%%%%%%%%%%%%%%%%%%%%%%%%%%%%%%%%%%%%%


\begin{thebibliography}{99}

\bibitem{MSSW}
J.~Madore, S.~Schraml, P.~Schupp and J.~Wess,
{\it Gauge theory on noncommutative spaces},
Eur. Phys. J. {\bf C}, in press,
hep-th/0001203.

\bibitem{CDS} A. Connes, M. R. Douglas, A. Schwarz,
\ti{\jhep}{9802}{003}{1998}{Noncommutative Geometry and Matrix Theory:
Compactification on Tori}, hep-th/9711162.

\bibitem{SW}
N.~Seiberg and E.~Witten,
{\it String theory and noncommutative geometry},
JHEP {\bf 9909} (1999) 032,
hep-th/9908142.

\bibitem{JS}
B.~Jur\v co, P.~Schupp,
{\it Noncommutative Yang-Mills from equivalence of star products},
Eur. Phys. J. {\bf C 14}, 367 (2000), hep-th/0001032.

\bibitem{JSW}
B.~Jur\v co, P.~Schupp and J.~Wess,
{\it Noncommutative gauge theory for Poisson manifolds},
Nucl. Phys. {\bf B}, in press,
hep-th/0005005.

\bibitem{JSW2}
B.~Jur\v co, P.~Schupp and J.~Wess,
{\it Nonabelian noncommutative gauge theory},
in preperation.

\bibitem{Weyl} 
H. Weyl, \ti{Z. Physik}{46}{1}{1927}{Quantenmechanik
und Gruppentheorie}; {\em The theory of groups and quantum
mechanics}, Dover, New-York (1931), translated from {\em
Gruppentheorie und Quantenmechanik}, Hirzel Verlag, Leipzig (1928).

\bibitem{Wigner} 
E. P. Wigner, \ti{\pr}{40}{749}{1932}{Quantum
corrections for thermodynamic equilibrium}.

\bibitem{Moyal} 
J. E. Moyal, \ti{Proc. Cambridge Phil.
Soc.}{45}{99}{1949}{Quantum mechanics as a statistical theory}.

\bibitem{BFFLS} F. Bayen, M. Flato, C. Fronsdal, A. Lichnerowicz, D.
Sternheimer, \ti{Ann. Physics}{111}{61}{1978}{Deformation theory and
quantization. I. Deformations of symplectic structures}.

\bibitem{Kontsevich} M. Kontsevitch, \emph{Deformation quantization of Poisson
manifolds, I,}\\q-alg/9709040.

\bibitem{Sternheimer} D. Sternheimer, \emph{Deformation Quantization: Twenty
Years After}, \\math/9809056.

\bibitem{WZ}
J.~Wess and B.~Zumino,
{\it Covariant differential calculus on the quatum hyperplane}, 
Nucl.\ Phys.\ Proc.\ Suppl.\  {\bf 18B} (1991) 302.


\bibitem{W} J.~Wess, {\it $q$-deformed Heisenberg Algebras}, in H.~Gausterer,
H.~Grosse and L.~Pittner, eds., Proceedings of the 38. Internationale
Universit\"atswochen f\"ur Kern- und Teilchenphysik, no.\ 543 in Lect. Notes
in Phys., Springer-Verlag, 2000, Schladming, January 1999, math-ph/9910013.

\bibitem{Wise} J.~Gomis, T.~Mehen and M.~B.~Wise, {\it Quantum Field Theories
With Compact Noncommutative Extra Dimensions}, hep-th/0006160.

\bibitem{Bonora}
L.~Bonora, M.~Schnabl, M.~M.~Sheikh-Jabbari and A.~Tomasiello,
{\it Noncommutative SO(n) and Sp(n) gauge theories},
hep-th/0006091.

\bibitem{Ho} P.-M.~Ho, {\it Twisted Bundle on Noncommutative Space and U(1)
Instanton}, hep-th/0003012.

\bibitem{AG} L.~Alvarez-Gaume, S.~R.~Wadia, {\it Gauge Theory on a 
Quantum Phase Space}, hep-th/0006219.


\end{thebibliography}
\end{document}